\documentclass{aa}
\usepackage{graphics}
\usepackage{natbib}
\usepackage{graphicx}
\usepackage{amssymb}
\begin{document}

\title{X-rays from $\alpha$ Centauri - The darkening of the solar twin}

\author{J. Robrade\inst{1}, J.H.M.M. Schmitt\inst{1}, and F. Favata\inst{2}}
\institute{
Hamburger Sternwarte, Universit\" at Hamburg, Gojenbergsweg 112,
D-21029 Hamburg, Germany
\and
Astrophysics Division -- Research and Science Support Department of ESA, ESTEC,
Postbus 299, NL-2200 AG Noordwijk, The Netherlands
}

\authorrunning{Robrade, Schmitt \& Favata}
\titlerunning{$\alpha$ Cen in X-rays}
\offprints{J. Robrade}
\mail{jrobrade@hs.uni-hamburg.de}
\date{Received 26 April 2005 / Accepted 11 July 2005}

\abstract{

We present first results from five XMM-Newton observations of the binary system $\alpha$ Centauri,
which has been observed in snapshot like exposures of roughly two hours each during the last 
two years. In all our observations the X-ray emission of the system is dominated 
by $\alpha$\,Cen B, a K1 star. 
The derived light curves of the individual components reveal variability on short timescales and a flare
was discovered on $\alpha$\,Cen B during one observation. 
A PSF fitting algorithm is applied to the event
distribution to determine the brightness of each component during the observations.
We perform a spectral analysis with multi-temperature models 
to calculate the X-ray luminosities.
We investigate long term variability and possible 
activity cycles of both stars and find
the optically brighter component $\alpha$\,Cen A,
a G2 star very similar to our Sun, to have fainted in X-rays by at least 
an order of magnitude during the observation program, a behaviour never observed before on
$\alpha$ Cen A, but rather similar to the X-ray behaviour observed with XMM-Newton on HD~81809.
We also compare our data with earlier spatially resolved observations performed over the last 25 years.

}
\maketitle

\section{Introduction}
\label{intro}

The visual binary system $\alpha$ Centauri AB (\object{HD~128620}/1) is the nearest stellar system 
consisting of a G2V (A) and a K1V (B) star at a distance of 1.3\,pc; in addition,
the M~dwarf Proxima Centauri is in common proper motion with the $\alpha$\,Cen\,A/B system.
The two components $\alpha$\,Cen A and B are separated by roughly 25 AU, with an orbital period of 80 years.
The age of the system is thought to be slightly larger than that of the Sun, correspondingly
both stars are also slow rotators (periods are 29\,(A) and 42\,(B) days) with a rather inactive corona. 
References and further literature can be found in \cite{letgs}, who analyzed {\it Chandra} LETGS data
from $\alpha$ Cen and in \cite{pag04}, who analysed UV data using HST/STIS. 
The latter authors determined an emission measure distribution of $\alpha$\,Cen\,A 
from UV to X-rays measurements and found it to be comparable to the quiet Sun, 
making the slightly more massive star $\alpha$\,Cen\,A a nearly perfect solar twin.
This finding immediately raises the question, whether an activity cycle as observed on the Sun, is also 
present on $\alpha$\,Cen.  While long-term measurements 
of chromospheric activity have been performed for several decades \citep{bal95}, 
clear indications for X-ray, i.e. coronal, activity 
cycles on other stars than the Sun were only recently found on a few objects.

\cite{hem03} analysed a time series of 4.5 years of ROSAT HRI data, taken at intervals of typically 6~months, 
on the stars 61~Cyg A and B with
well determined chromospheric cycles of 7 and 12 years respectively, and find that coronal cycles are the
dominant source of long-term X-ray variability for both stars. 
61~Cyg was monitored also with XMM-Newton over the last years
and results of these observations will be presented in a forthcoming paper.
Similarly, \cite{fav04} obtained
a time series of 2.5 years of XMM-Newton data, again obtained at 6 month intervals, of the G~star
HD~81809, which has a pronounced cycle of 8.2 years, and find clear evidence for large amplitude 
X-ray variability in phase with the known (chromospheric) activity cycle.

The $\alpha$\,Cen system has been studied before in X-rays with several missions, 
e.g. {\it Einstein}, ROSAT, ASCA and recently {\it Chandra}.
The two components were already spatially separated with {\it Einstein} \citep{eins}.
The K~star was found to dominate the X-ray emission and is usually a factor 2--3 brighter 
than the G~star at typical
energies above 0.2\,keV in previous observations. 
In two ROSAT HRI monitoring campaigns, performed in 1996 
with nearly daily measurement for a month each, light curves for the individual components were obtained and
indications for flaring and a decrease in X-ray brightness of 30\,\% over 20 days, which could be due to 
rotational modulation, were found on $\alpha$\,Cen B \citep{schmitt04}. 
Separate high resolution spectra were first obtained with the {\it Chandra} LETGS \citep{letgs}, which
revealed solar like properties for both stars, e.g. the FIP effect and an
emission measure distribution dominated by cool plasma with temperatures of 1-3\,MK.
The K~star was found to be slightly hotter and dominates the emission measure above 1.5\,MK, while at lower 
temperatures the G~star is the stronger component. 

In order to study possible coronal activity cycles of solar-like stars we initiated a long-term monitoring
program of a small number of objects and first results on HD~81809 were presented by \cite{fav04}.   
Within the context of this ongoing monitoring program
the $\alpha$ Cen system was repeatedly observed with XMM-Newton, 
and here we report first results on the  $\alpha$ Cen system. 
In Sect.\,\ref{obsana} we describe the observations and the methods used for data analysis.
In Sect.\,\ref{results} we present the results subdivided into different physical topics, in Sect.\,\ref{dis}
we discuss the findings in comparison with previous observations followed by our conclusions in Sect.\,\ref{summ}.

\section{Observations and data analysis}
\label{obsana}

The target $\alpha$\,Cen was repeatedly observed with XMM-Newton using almost identical detector setups and with
exposure times in the range of 5\,-\,9\,ksec. We present data from five observations separated
by roughly half a year each, which allows us to study short time behaviour during individual exposures
as well as long term variations on timescales of several month up to years.
Useful data were collected in all X-ray detectors, which were operated simultaneously onboard XMM-Newton, 
respectively the EPIC (European Photon Imaging Camera), consisting of the MOS and PN detectors
and the RGS (Reflection Grating Spectrometer).
The MOS and PN observations were performed in the small and large window mode with the thick filter.
The OM (Optical Monitor) was blocked due to the brightness of the target.

A description of the observations is provided in Table~\ref{obs} and
a detailed description of the XMM-Newton instruments can be found in \cite{xmm}.

\begin{table}[!ht]
\caption{\label{obs}Observation log of $\alpha$\,Cen, MOS1}
{\scriptsize
\begin{tabular}{lccc}\hline\hline
Obs.ID & Obs.Mode &  Obs. Time  & Dur. (s)\\\hline
0045340901 & SW/thick & 2003-03-04T13:50-15:47 & 6850 \\
0045341001 & SW/thick & 2003-09-15T14:55-17:04 & 7560 \\
0045341101 & SW/thick & 2004-01-29T14:16-15:45 & 5160 \\
0045340401 & SW/thick & 2004-07-29T03:48-05:59 & 7660 \\
0143630501 & LW/thick & 2005-02-02T14:47-17:13 & 8770 \\\hline
\end{tabular}
}
\end{table}

The data were reduced with the standard XMM-Newton Science Analysis System (SAS)
software, version 6.0. 
Images, light curves and spectra were produced with standard
SAS tools and standard selection criteria were applied for filtering the data, see \cite{sas}.
Spectral analysis was carried out with XSPEC V11.3 \citep{xspec}.

\begin{figure}[!ht]
\includegraphics[width=85mm]{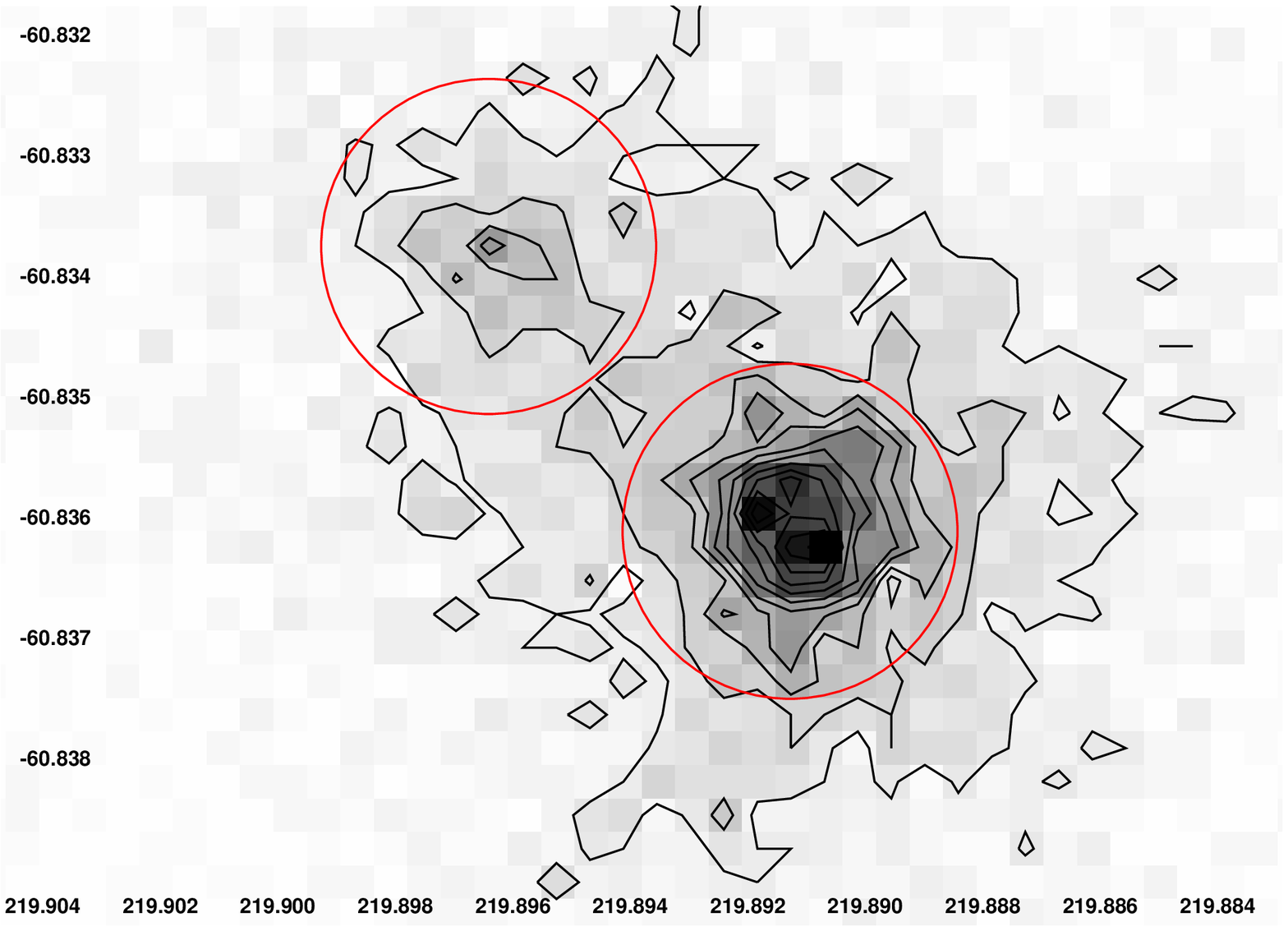}
\includegraphics[width=85mm]{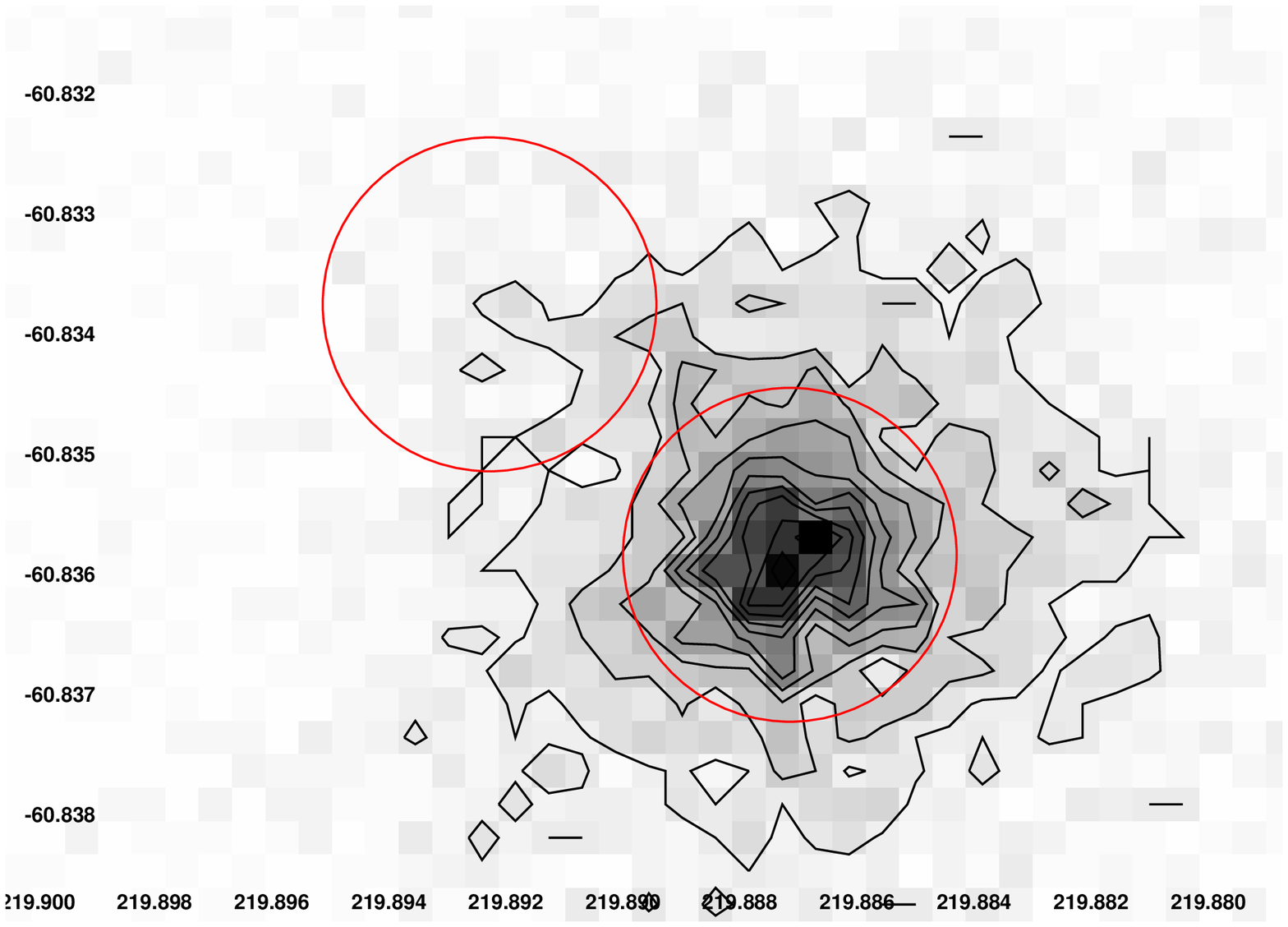}
\caption{\label{image}Image of the system obtained with MOS1 during the March 03 (top) and
February 05 (bottom) exposure with contours and used extraction regions (red) for the two components overlayed. 
Image creation is identical and the counts per image are very similar, 
the darkening of the  $\alpha$\,Cen\,A component at the upper left is striking.}
\end{figure}

For imaging we use data taken with the MOS1 detector, which has a more spherical symmetric PSF shape 
compared to MOS2 and a better spatial resolution than the PN detector.
Spectral analysis of EPIC data is performed in the energy band between 0.2\,-\,5.0\,keV, but sufficient signal
is mostly present only up to $\sim$\,2\,keV. For the RGS first order spectra in the full energy range, 
i.e. 0.35\,-\,2.5\,keV (5\,-\,35\,\AA), are used. 
While the RGS has the highest spectral resolution,
the EPIC detectors are able to measure higher energy X-rays with higher sensitivity, 
with the MOS detectors providing a slightly better spectral resolution and the PN detector
providing greater sensitivity.
Data of the same detector type, i.e. RGS1 and RGS2 and MOS1 and MOS2, 
were analyzed simultaneously but not co-added. Data quality is generally good, only
the MOS2 data of the July 04 observation are corrupted and had to be discarded from further analysis.
The background was taken from source free regions on the detectors. 
Our fit procedure is based on $\chi^2$ minimization, therefore spectra 
are rebinned to satisfy the statistical demand of a minimum value of 15 counts per spectral bin.

For the analysis of the X-ray spectra we use multi-temperature models with variable but tied abundances,
i.e. the same abundance pattern in all temperature components.
Such models assume the emission spectrum of a collisionally-ionized optically-thin gas 
as calculated with the APEC code, see e.g. \cite{apec}. 
Abundances are calculated relative to solar photospheric values as given by 
\cite{and89}. For iron and oxygen we use the values of \cite{grev98}.
The applied model uses two temperature components,
models with additional temperature components were checked, but did not improve the fit 
results significantly.
Due to the lower spectral resolution of the EPIC detectors, for those elements where features are 
most prominent only in the RGS,
the RGS values were taken, for elements without clearly recognizable lines, e.g. Al, Ca, Ni, solar 
values were used.
X-ray luminosities were then calculated from the resulting best fit models.
Due to the proximity of the stars absorption in the interstellar medium is negligible at 
the wavelengths of interest and was not applied in our modelling.

\section{Results}
\label{results}
\subsection{Investigation of images and light curves}
\label{lcana}

In Fig.\ref{image} we show two images of the system taken with the MOS1 detector during the first (March 03) and last
(Feb. 05) exposure of our data sample. Significant changes in the luminosity of the components are obvious,
especially the strong dimming of $\alpha$\,Cen\,A.
To investigate these changes we use individual light curves and a PSF fitting algorithm, which is applied to the
event distribution in the sky plane.

\begin{figure}[!ht]

\includegraphics[width=88mm,height=44mm]{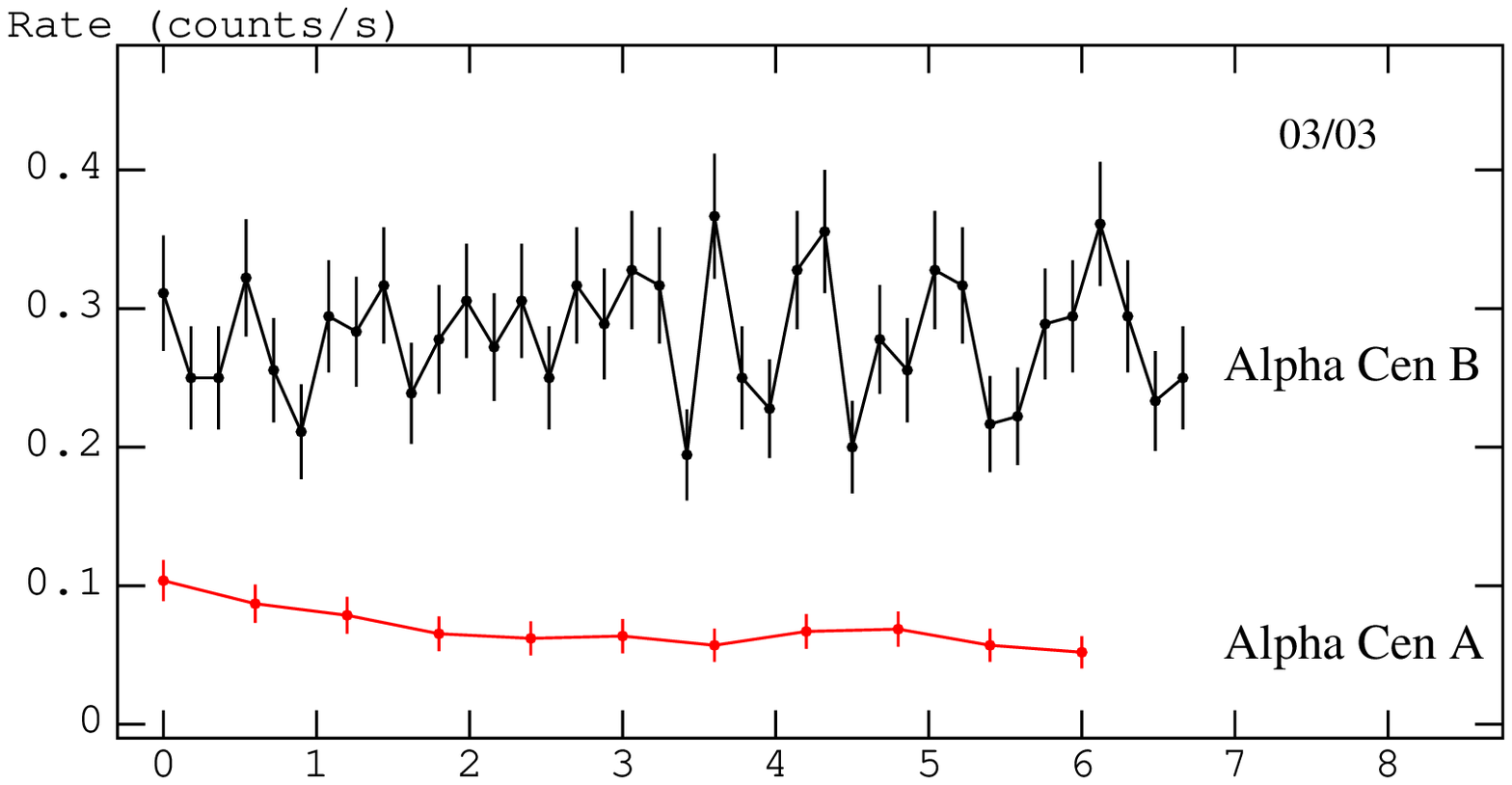}

\vspace*{-5.3mm}
\includegraphics[width=88mm,height=44mm]{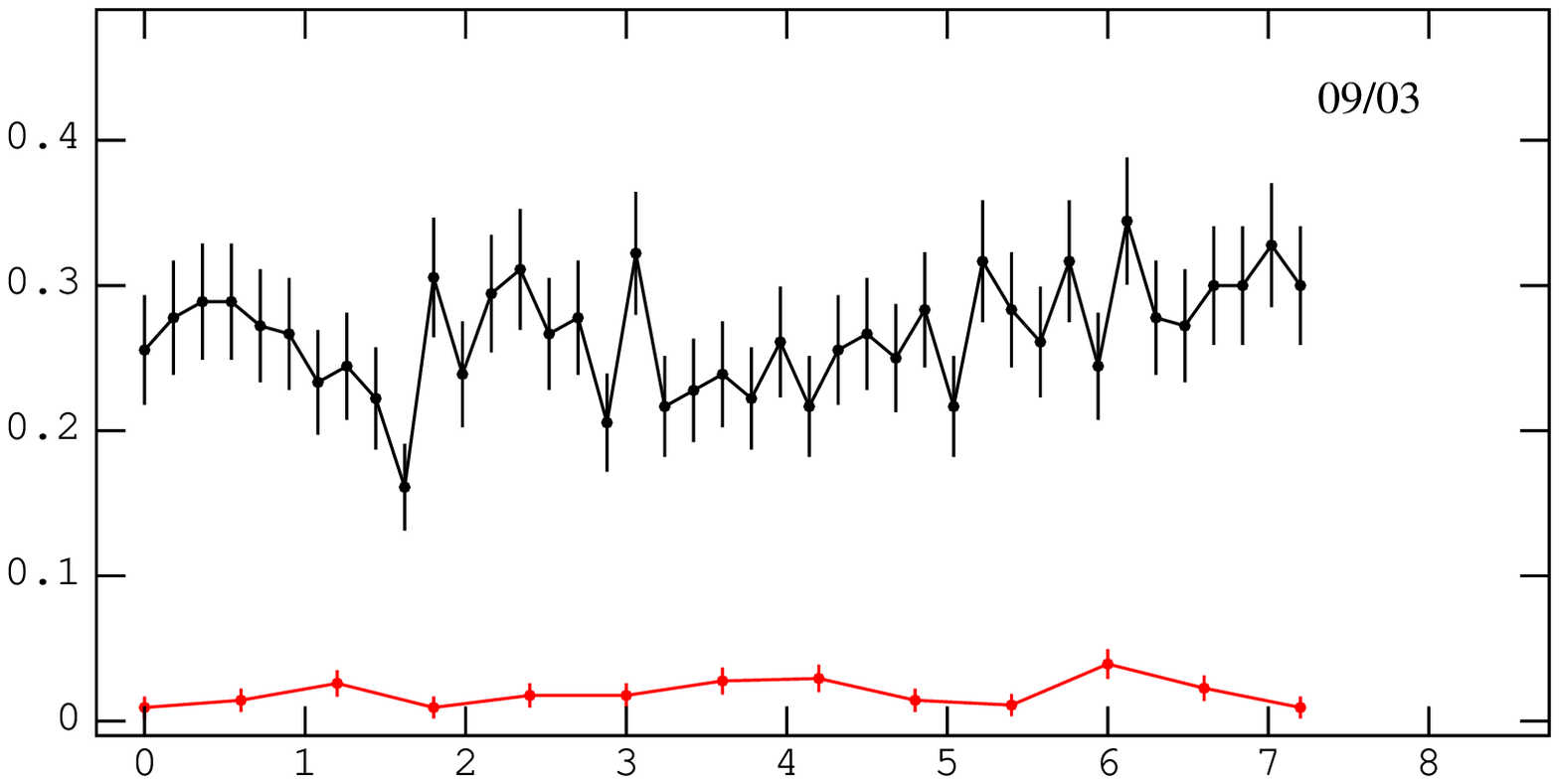}

\vspace*{-5.2mm}
\includegraphics[width=88mm,height=44mm]{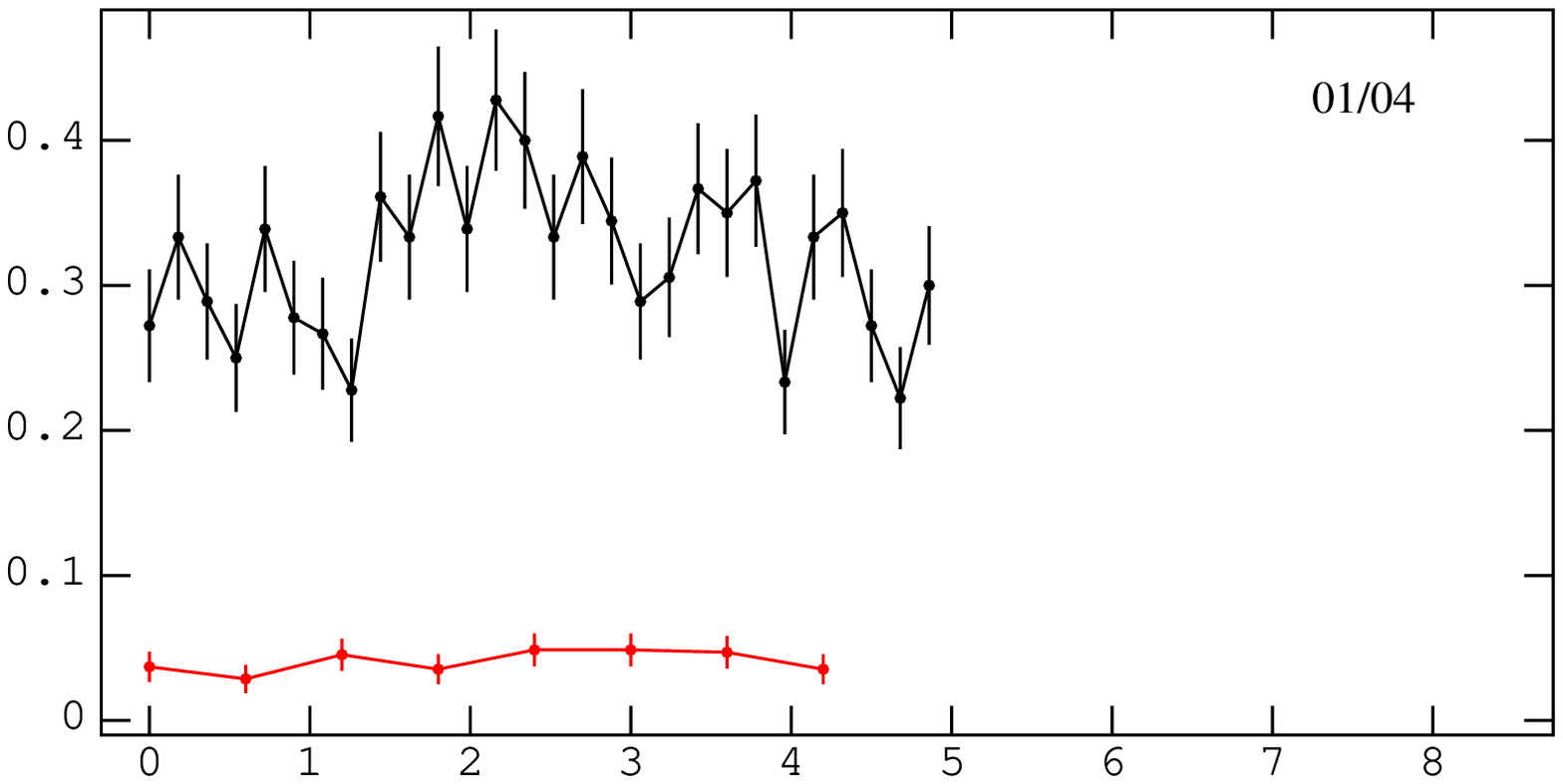}

\vspace*{-5.5mm}
\includegraphics[width=88mm,height=44mm]{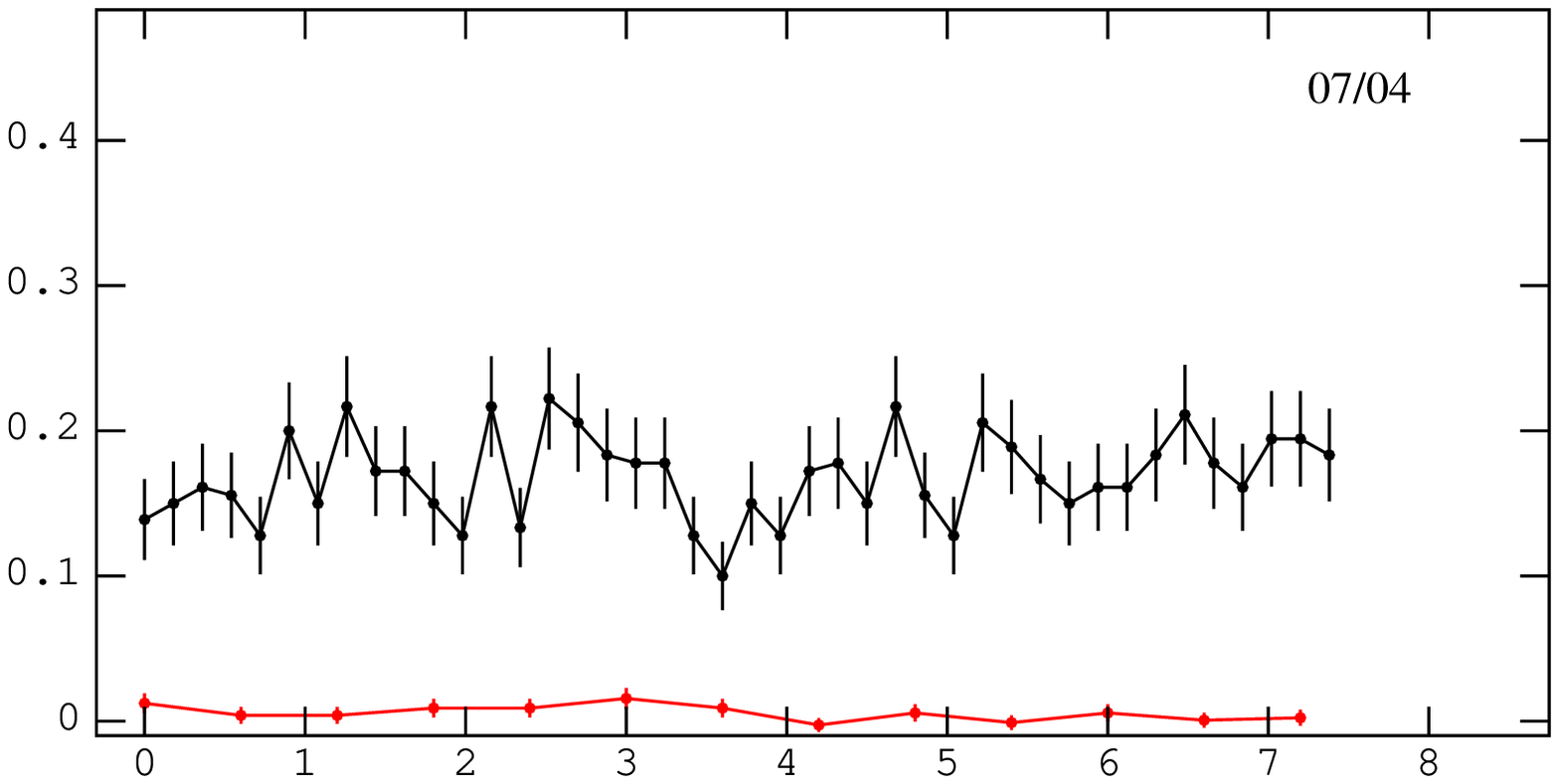}

\vspace*{-5.2mm}
\includegraphics[width=88mm,height=44mm]{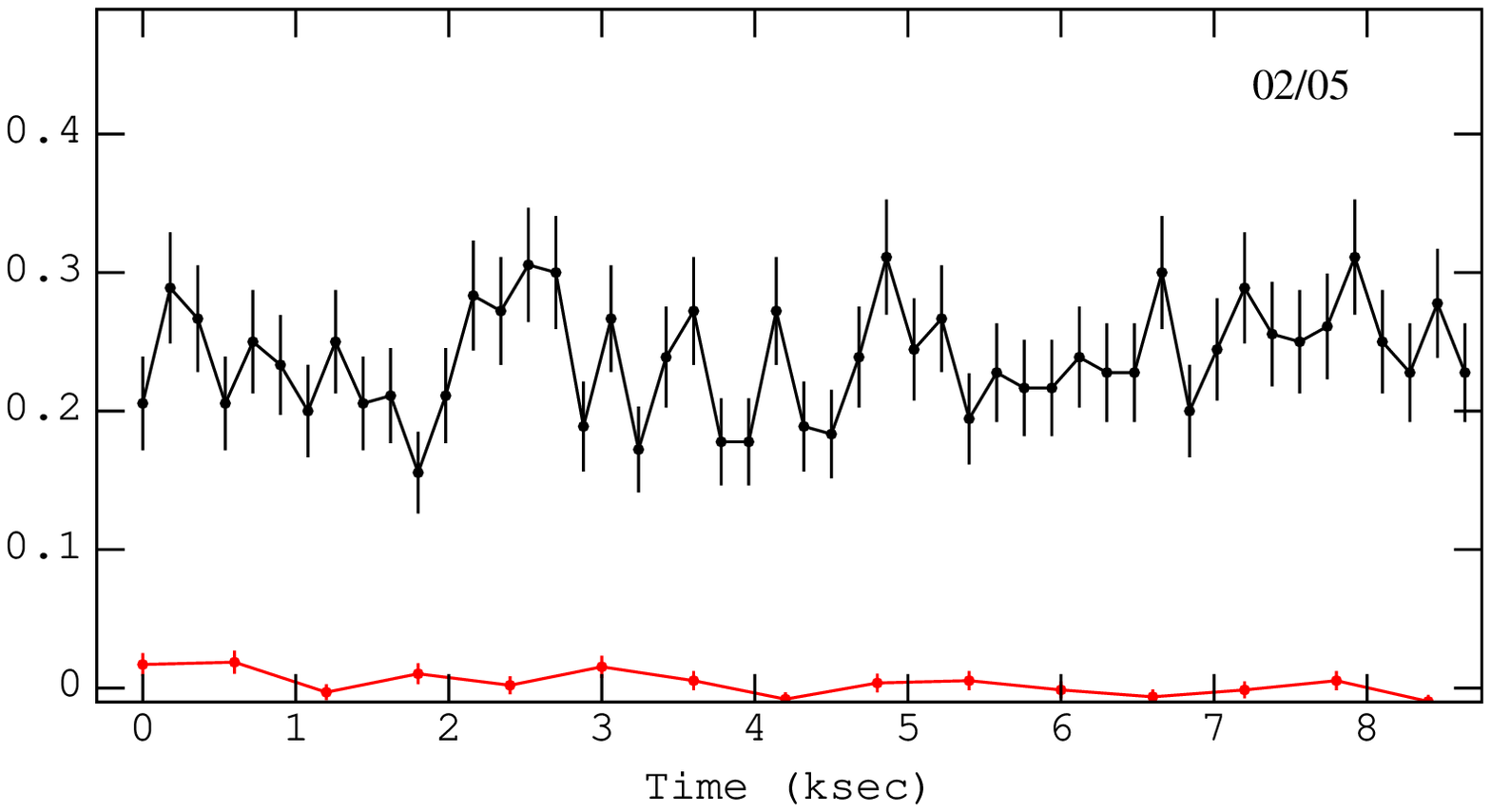}

\caption{\label{lcs}Light curves of $\alpha$\,Cen A/B (red/black) in chronological order as extracted from MOS1 data
with 10/3\,min. binning. The plot sizes are identical for all observations to allow better comparison.}
\end{figure}

The MOS1 light curves of $\alpha$\,Cen\,A and $\alpha$\,Cen\,B for the five observations, 
separated roughly half a year each, are shown in Fig.\,\ref{lcs}.
The light curves were extracted from a circle with 5\,\arcsec\,radius
around the respective position of the sources, the temporal binning is 600/180\,s for the A/B component. 
For $\alpha$\,Cen\,A
the measured count rate in a region mirrored at the position of $\alpha$\,Cen\,B is subtracted
to account for contamination through the
much brighter B component; further background contributions are negligible.
It is obvious that $\alpha$\,Cen\,B is always the brighter X-ray source, but the count ratio A/B differs significantly.
Looking at the individual light curves, $\alpha$\,Cen\,A is mainly constant, only in March 03 a steady decline is
visible throughout the observation.
$\alpha$\,Cen\,B exhibits short time variability in all observations and a small flare 
occurred during the Jan. 04 exposure. 
The mean count rate of $\alpha$\,Cen\,B is 
comparable during the first three observations, roughly 50\% lower in the fourth one 
and has nearly recovered in the Feb. 05
exposure, while $\alpha$\,Cen\,A has declined by more than an order of magnitude over the two years. 

To quantify the individual count rates more precisely, we
fitted two instrumental PSFs to the event distribution in the sky-plane taken from a 50$\times$50\,\arcsec\,box
around the position of $\alpha$\,Cen following the procedure described by \cite{eqpeg}. 
After adjusting the PSF shape we kept these fixed, fitted the position of the sources
and derived the counts for each of the two components.
Small variations of the position were allowed to account for the small proper motion and 
the two components are 
detected in all observations very close to their calculated positions, only for the February 05 observation 
the position of the $\alpha$\,Cen\,A had to be fixed because of the weakness of the source at that time. 
For this most critical observation we derive values for the count ratio B/A in the range
65--85, depending on the exact position of the sources. 
This error is of the same magnitude as if assuming Poissonian errors for the
derived counts of each source.

\begin{table}[!ht]
\setlength\tabcolsep{3.5pt}
\caption{\label{tabresl}PSF fit results. Poissonian errors are calculated for the derived ratio.}
{\scriptsize
\begin{tabular}{lccccc}\hline\hline
Obs.& Mar. 03 & Sept. 03 & Jan. 04 & July 04 & Feb. 05\\\hline
Live-time(s)& 6646 & 7315& 4952 &7423 &8661 \\
Cts (total) & 6152 & 5775& 4690 & 3477 & 5759\\
Cts\,($\alpha$CenB)&4806 & 5392 & 4161 & 3272 & 5694\\
Cts\,($\alpha$CenA)&1346 & 383& 529 & 205 & 76\\
Ratio B/A & 3.6\,$\pm$\,0.1 & 14.1\,$\pm$\,0.8& 7.9\,$\pm$\,0.4 & 16.0\,$\pm$\,1.2& 75.0\,$\pm$\,10.0\\\hline
\end{tabular}
}
\end{table}

Results of the PSF fitting procedure are presented in Table\,\ref{tabresl} and
are also in good agreement with estimates made from the individual light curves derived above. 
While $\alpha$\,Cen\,B exhibits a slowly varying brightness, only 
in July 04 it was significantly darker compared to the other observations, $\alpha$\,Cen\,A is mainly 
fading away throughout the campaign.

\subsection{Spectral analysis}
\label{specana}

Due to the low signal to noise of the data from the individual observations we modelled the data from 
the five observations
simultaneously, i.e. we neglect possible changes in the elemental abundances, 
while keeping only
the temperatures and emission measures as free parameters to account for the different flux levels and possible
coronal heating due to flaring. Because of overlapping PSFs for the individual components, the spectral analysis is 
performed on the $\alpha$\,Cen system as a whole.
However, individual fits of spectra taken from small extraction regions around the respective component
lead to comparable results for both components.
In this work the spectral models are primarily used to determine the X-ray flux for the individual sources.
We find that the derived fluxes and therefore luminosities vary only moderately within the
different applied models, introducing only
a small effect compared to differences arising from calibration uncertainties of the different detectors and
errors on the derived count rates.

\begin{figure}[!ht]

\vspace*{-2.0cm}
\includegraphics[width=90mm]{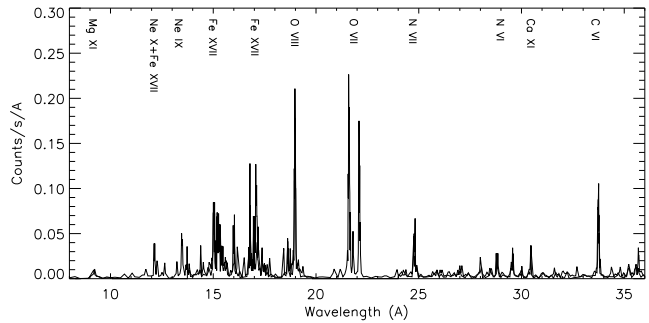}
\caption{\label{rgs}The RGS spectrum of the Feb. 05 exposure, an almost pure spectrum of $\alpha$\,Cen\,B.}
\end{figure}

The RGS spectrum shown in Fig.\,\ref{rgs} is dominated by emission lines with no strong continuum visible.
The most prominent features are labeled. These lines are
formed at relatively cool temperatures, plasma with
temperatures in the range of 1-4\,MK provides here the dominant contributions to line formation.

\begin{figure}[!ht]
\includegraphics[width=50mm,height=85mm,,angle=-90]{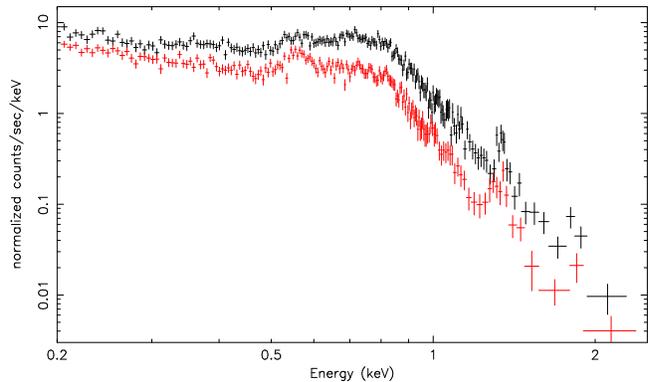}
\caption{\label{pn}Spectra of the $\alpha$\,Cen system obtained with the PN detector: 
03/03 (black/top), 07/04 (red/bottom).
Despite the different contributions from each component and absolute luminosities, both spectra are very similar
and soften only slightly with decreasing flux.}
\end{figure}

The PN spectra of two selected observations are shown in Fig.\,\ref{pn}. 
At energies above 1.0\,keV the here non-resolved He-like triplets of magnesium (at 1.35\,keV) and silicon (at 1.85\,keV)
are visible, representing the 'hottest' features in the spectra of these stars. The \ion{Mg}{xi} and \ion{Si}{xiii} lines
form at temperatures of 6\,--\,10\,MK and are more prominent during the brighter exposures or during flaring. 
Overall spectral changes are also a little more pronounced at higher energies. 
However, the general spectral shape is very similar for the exposures despite the different flux levels and the 
different contributions of $\alpha$\,Cen\,A to the flux of the system.

To determine the spectral properties and to calculate the X-ray luminosities of the components, we fitted the spectra of the
different instruments with multi-temperature spectral models as described in Sect.\ref{obsana}.
We find a two temperature fit to be sufficient, only during the flare-observation the model improves 
with a third hotter (8-10\,MK) component, which contributes around 1-2\,\% to the emission measure of the $\alpha$\,Cen\,system.
Since we are not able to clearly separate the spectra of the individual components 
and $\alpha$\,Cen\,B is the far brighter X-ray source, we consider all spectral results
presented here strongly dominated by $\alpha$\,Cen\,B, especially for the later observations where $\alpha$\,Cen\,A has darkened.  
A spatially resolved spectral analysis of the two components using data of the {\it Chandra} LETGS \citep{letgs},
where the sources are clearly separated due to the better spatial resolution,
revealed solar like properties for both stars with a slightly hotter corona to be present in $\alpha$\,Cen\,B.
Although a detailed spectral analysis is beyond the scope of this work, we would like to point out that consistent results
were obtained within our modelling. All X-ray luminosities are given in the 0.2\,--\,2.0\,keV band unless otherwise indicated.

\begin{table}[!ht]
\caption{\label{specres}X-ray luminosity in $10^{27}$\,erg/s of the $\alpha$\,Centauri system in the 0.2-2.0\,keV band
as derived with the different detectors, for the MOS also separated into three energy bands
(low: 0.2-0.5, medium: 0.5-0.75, high: 0.75-2.0 keV).}
{\scriptsize
\begin{tabular}{lrcc}\hline\hline
Obs.Date & $L_{\rm X}$ -- MOS - (low/med./high) & RGS &PN\\\hline
Mar. 03 & 2.47 - (1.42\,/\,0.64\,/\,0.42)& 2.30 &2.86\\
Sept. 03 & 2.17 - (1.27\,/0.55\,/\,0.35)& 2.02 &2.60 \\
Jan. 04 & 2.52 - (1.41\,/0.64\,/\,0.47) & 2.37 &2.86 \\
July 04 & 1.31 - (0.79\,/\,0.32\,/\,0.19) & 1.19 &1.79 \\
Feb. 05& 1.86 - (1.11\,/\,0.47\,/\,0.28) & 1.62 &2.25 \\\hline

\end{tabular}
}
\end{table}

In Table\,\ref{specres} we show the derived X-ray luminosities of the $\alpha$\,Centauri system.
The derived fluxes for the individual instruments onboard of XMM-Newton differ slightly, 
with the MOS results taking an intermediate place between RGS and PN. The PN models predict higher fluxes, especially
at lower energies. The subdivision of the MOS results into three energy band exhibits that flux changes
are strongest above 0.75\,keV and they become more pronounced at even higher energies; 
but since there is actually not much emission at these energies, the decline in X-ray brightness
has to be attributed mainly to a decrease of the emission measure instead of a cooling. 
This is also reflected by
the average coronal temperatures, which are around 2.8\,MK and differ only by 0.1\,MK between the different exposures.

\subsection{A Flare on the K star}
\label{flare}

The individual MOS1 light curves shown in Fig.\,\ref{lcs} clearly identify $\alpha$\,Cen\,B 
as the flaring star in the January 2004 observation,
which confirms the nature of the K star as a flare star, as already suggested by \cite{schmitt04}.
Due to the greater sensitivity we use the PN data of the total system to investigate details of the flare.
The luminosity of $\alpha$\,Cen\,A is much lower than that of
$\alpha$\,Cen\,B and is nearly constant over the whole exposure; it therefore induces only an offset on the light curves. 

\begin{figure}[!ht]
\includegraphics[width=90mm]{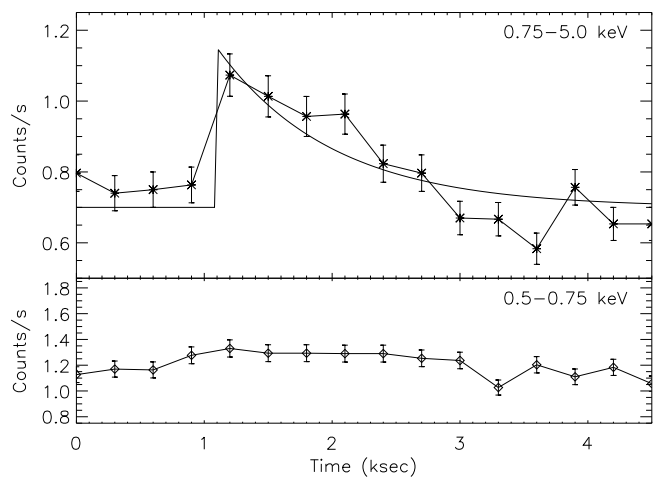}
\caption{\label{lcf}Light curve of the system in two energy bands with the 
flare on $\alpha$\,Cen\,B that occurred on January, 29, 2004, 
PN data with 300\,s binning. Overlayed on the hard band is a flare model assuming an exponential decay.}
\end{figure}

The PN light curves from this observation of the system in two energy bands (medium: 0.5\,--\,0.75\,keV, hard: 0.75\,--\,5.0\,keV)
are shown in Fig.\,\ref{lcf}. In the soft band, i.e. below 0.5\,keV, the flare is hardly visible.
Overlayed on the hard band is a simple exponential flare model of the form \hbox{$c(t)=C_{0}+A\,e^{(-t/\tau)}, t>0$}
with an exponential decay time of $\tau$=900\,s. 
The flare is much more pronounced in the hard band, a typical behaviour of stellar flares. 
Using the average energy per photon as derived from the spectral models in Sect.\ref{specana}
we calculate the energy released by this flare event and
we derive a total flare energy release of $\sim 5\times10^{29}$\,erg above 0.5\,keV with 60\% percent 
measured in the hard and 40\% percent in the medium band. Therefore this flare is comparable to a typical solar flare,
where the energy release is in the order of $10^{29}$\,erg while the largest solar flares release up to $10^{32}$\,erg.

\subsection{Long term behaviour}
\label{long}

\begin{figure}[!ht]
\vspace*{-1.5cm}
\includegraphics[width=90mm]{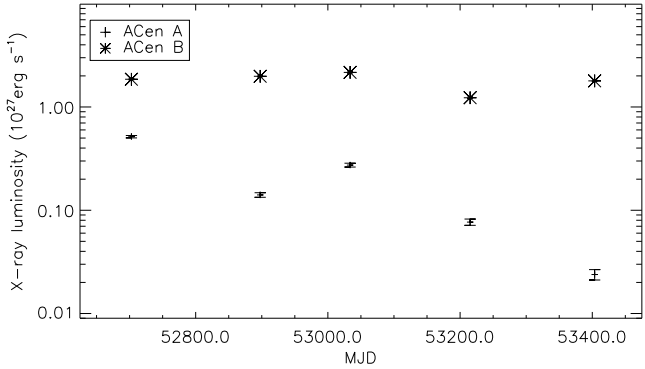}
\caption{\label{lcall}Long term light curve of $\alpha$\,Cen\,A and B for the five XMM-Newton observations. 
Errors are Poissonian errors on the derived counts only.}
\end{figure}

In Fig.\,\ref{lcall} we show the calculated 0.2\,--\,2.0 keV X-ray luminosities
of $\alpha$\,Cen\,A and B as derived from spectral models applied to MOS1 data 
combined with the results from the PSF fitting procedure; the numerical values
are given in Table\,\ref{fluxres}.  In this context
it is important to note that individual fluxes are calculated from the 
measured counts of the individual components with a model derived for the sum of both components.
In the considered energy band the average energy per photon is slightly lower for the G~star, so especially
in case of activity on $\alpha$\,Cen\,B the flux of $\alpha$\,Cen\,A may actually be somewhat overestimated.

\begin{table}[!ht]
\caption{\label{fluxres}Derived X-ray luminosity in $10^{27}$\,erg/s for the two components $\alpha$\,Cen A and B from MOS1 data.}
{\scriptsize
\begin{tabular}{lccccc}\hline\hline
Obs.& Mar. 03 & Sept. 03 & Jan. 04 & July 04 & Feb. 05\\\hline
$L_{\rm X}$ (A) & 0.52 & 0.14 & 0.27 & 0.08 & 0.02\\
$L_{\rm X}$ (B) & 1.86 & 1.98 & 2.16 & 1.22& 1.79 \\\hline
\end{tabular}
}
\end{table}

The X-ray emission of $\alpha$\,Cen\,B is variable within a factor of two also in the 
absence of stronger flares.  However,
there is no strong trend visible over the whole observation campaign, only the  
July 04 observation stands out, exhibiting the lowest X-ray luminosity for $\alpha$\,Cen\,B.
The most remarkable result is that $\alpha$\,Cen\,A, a nearly solar twin, is found to be strongly
variable by a factor of 10\,--\,20. Such a behaviour has to our knowledge never been observed before on this star.

\section{Discussion}
\label{dis}

Although X-ray luminosities are subject to the usual calibration uncertainties of different instruments and
although flux ratios do depend on the selected energy range, we are able to compare our measurements with 
X-ray data taken over the last 25 years.
Since the X-ray spectra of both components decline very steeply towards higher energies, there is very little 
flux above 2.0\,keV, and thus only the lower energy threshold matters for a calculation of a broad band X-ray 
luminosity. To give an example, in our model we find that the derived energy flux increases by $\sim$\,25\% if measured
above 0.15\,keV instead of 0.2\,keV.

In 1979 a {\it Einstein} HRI (0.15\,--\,3.0 keV) observation \citep{eins} yielded X-ray 
luminosities of $1.2/2.8\times10^{27}$\,erg\,s$^{-1}$
for $\alpha$\,Cen\,A/B, resulting in a B/A ratio of 2.3. 
A large number of individually short ROSAT HRI observations ($\sim$\,40)
taken in 1996 indicated a B/A ratio varying between 2.0\,--\,3.5, excluding a probable
flare event observed on $\alpha$\,Cen\,B; X-ray luminosities (0.2\,--\,2.0 keV)
were about $1.0-1.3/2.6-3.6\times10^{27}$\,erg\,s$^{-1}$ for $\alpha$\,Cen\,A/B \citep{schmitt04}.
The more recent {\it Chandra} LETGS exposure \citep{letgs} taken in 1999 yielded
luminosities for $\alpha$\,Cen\,A/B of $0.9/1.6\times10^{27}$\,erg\,s$^{-1}$ in the energy range 0.15\,--\,4.0 keV 
with a B/A ratio of 1.9.
No variability of $\alpha$\,Cen was reported for an {\it Einstein} IPC (4.5\,h) and the {\it Chandra} (22.5\,h) observation.

Only the March 2003 exposure of our XMM-Newton campaign has exhibited a comparable B/A ratio; in this observation
we derive X-ray luminosities of $0.5/1.9\times10^{27}$\,erg\,s$^{-1}$ for $\alpha$\,Cen\,A/B in the 0.2\,--\,2.0 keV band.
The X-ray luminosity of $\alpha$\,Cen\,B is thus comparable with previous measurements  
and no variations larger than a factor of two, 
which are already present within the different exposures of the XMM-Newton campaign,
are found over a time interval of roughly 25 years. 
These findings support the scenario of a stable corona, where variations of the emission 
can be explained by a long-term activity cycle as indicated by emission line variability in
IUE data \citep{ayres95} covering roughly 10 years
and short term activity in small surface areas that are additionally subject to rotational modulation.
The period of this cycle, however,  has to be long, probably ten or more years,
while the relative modulation of the overall X-ray luminosity has to be much smaller than on the Sun to be
consistent with the X-ray data.

But what about $\alpha$\,Cen\,A? 
Although moderate short term variability also seems to be present, no indications
were found for a long-term activity cycle so far. 
Comparing the derived X-ray luminosities, the values from the 1979\,--\,1999 observations differ
by no more than 50\% with the lowest value measured in 1999. 
With the March 03 exposure being already another 40\% below the 1999 value, the
X-ray output drops by more than a magnitude within two years. 
In the February 05 exposure (cf., Fig. \ref{image}) we can hardly recognise an X-ray binary at all, and the
derived flux for $\alpha$\,Cen\,A has declined to $2.5\times10^{25}$\,erg\,s$^{-1}$. 
We do point out that the decline in X-ray luminosity observed with XMM-Newton cannot be explained 
by a pure temperature effect.
The assumption, that all of the X-ray emitting plasma has a temperature of only 1\,MK with constant emission measure, 
results in 
a decrease of X-ray luminosity by a factor of two and is thus not sufficient to explain our XMM-Newton observations.
Rather, a strong decrease of the total emission measure is necessary to explain our findings. 
While smaller differences in the long term evolution of X-ray luminosity may be
explained by the use of the various instruments, the decline seen over the XMM-Newton campaign can only be explained by 
a X-ray activity cycle or an irregular event. While no definite statement can be made about an irregular event,
the scenario of an activity cycle would require, that all previous X-ray measurements were made when $\alpha$\,Cen\,A was
near the 'high state' of its cycle. Putting all the observation dates together,
this would require a cycle with a duration of $\sim$\,3.4 years from maximum to maximum.
While chromospheric activity cycles on late-type stars were frequently found 
in \ion{Ca}{ii} H and K emission lines (Mt. Wilson S index) and periods
of a few years are not uncommon \citep{bal95}, the $\alpha$\,Centauri system was not observed in these programs due to its
location in the southern sky. 
Further, the long term variability studies of IUE-UV lines mentioned above found no
evidence for an activity cycle on $\alpha$\,Cen\,A although some scatter in the data is present.

In the X-ray regime indications for coronal activity cycles were found in three other stars. 
\cite{hem03} found evidence for coronal activity cycles in in both components of the K~dwarf binary 61 Cygni;
using ROSAT HRI data they determined X-ray luminosities for 61~Cygni A ($L_{\rm X}=1-3\times10^{27}$\,erg\,s$^{-1}$) and
61~Cygni B ($L_{\rm X}=0.4-1\times10^{27}$\,erg\,s$^{-1}$)
that correlate well with the chromospheric activity as measured in the \ion{Ca}{ii} H+K, and 
recently, \cite{fav04} presented an analysis of  the XMM-Newton data of the somewhat more active G2 
star HD~81809 ($L_{\rm X}=2-18\times10^{28}$\,erg\,s$^{-1}$), 
which also shows a drop in X-ray flux by more than an order of magnitude correlated with the Ca S index.
On $\alpha$\,Cen\,A long term X-ray variability is definitely present at a significant level, however the absence of 
comparative chromospheric activity data does not
allow a correlation analysis and therefore only future observations will allow to check if an activity cycle 
is present on $\alpha$\,Cen\,A, which
would then be the first X-ray activity cycle on a true solar analog.

\section{Conclusions}
\label{summ}
 
We have analysed five XMM-Newton observations of the $\alpha$\,Centauri system regularly performed over two years 
and determined light curves and fluxes for the $\alpha$\,Cen\,A (G2V) and B (K1V) components, which enables us to 
study short term behaviour as well as long term variability of the stellar activity of this system. 
The X-ray properties of both stars are characterized by a rather cool and inactive corona, but
the system is found to be strongly dominated by the K star, $\alpha$\,Cen\,B.
The X-ray luminosity of $\alpha$\,Cen\,B appears be variable within a factor of two and we are able
to confirm its nature as a flare star. The observed flare on $\alpha$\,Cen\,B
is probably one of the weakest stellar flare events, where
typical flare signatures like a well defined decay time were actually detected. 
A long term X-ray activity cycle on $\alpha$ Cen B, if present at all, has to be characterized by
a long period and/or a small modulation.

For $\alpha$\,Cen\,A we find a strong decline in X-ray luminosity by no less than an order of magnitude over 
the time span of our observations of two years,
a behaviour that was never observed before on this star during observations performed over the last 25 years.
This might then indicate a coronal activity cycle with all other observations having occurred - by chance -  
near the 'high state' or - alternatively - an irregular event. 
The absence of long term chromospheric activity data for these
stars make a definite statement on this point impossible. 
The observed trend of $\alpha$ Cen's X-ray luminosity is actually comparable to solar activity parameters as observed
with the Yohkoh SXT in the 0.3\,--\,3.0 keV band. \cite{acton} studied changes of the solar X-ray emission from 
1991 (near solar maximum) to 1995 (near solar minimum) and found that the average coronal temperature changed
only by a factor of 1.5, i.e. from 3.3\,MK to 1.9\,MK, while emission measure declined by a factor larger
than ten. A similar scenario would clearly explain the observed decline of $\alpha$\,Cen\,A's X-ray brightness.
The program is ongoing and future observations are scheduled
to enlightened the nature of the coronal variability of our Sun's neighbouring twin.
 
\begin{acknowledgements}
This work is based on observations obtained with XMM-Newton, an ESA science
mission with instruments and contributions directly funded by ESA Member
States and the USA (NASA).\\
This research has made use of the SIMBAD database, operated at CDS, Strasbourg, France.
(http://simbad.u-strasbg.fr)\\
J.R. acknowledges support from DLR under 50OR0105.

\end{acknowledgements}


\begin{thebibliography}{}
\bibitem[Acton(1996)]{acton}Acton, L.W. 1996, ASP Conf. Series, 109, 45
\bibitem[Anders \& Grevesse(1989)]{and89}Anders E., Grevesse N. 1989, Geo- et Cosmochimica Acta, 53, 197
\bibitem[Arnaud(1996)]{xspec}Arnaud, K.A. 1996, ASP Conf. Series, 101, 17
\bibitem[Ayres et al.(1995)]{ayres95}Ayres, T.R., Fleming, T.A., Simon, T., et al. 1995, ApJS, 96, 223
\bibitem[Baliunas et al.(1995)]{bal95}Baliunas, S.L., Donahue, R.A., Soon, W.H., et al. 1995, ApJ, 438, 269
\bibitem[Ehle et al.(2003)]{xmm}Ehle, M., Breitfellner, M., Gonzales Riestra, M., et al. 2003, XMM-Newton User's Handbook
\bibitem[Ehle et al.(2004)]{sas}Ehle, M., Pollock, A.M.T., Talavera, A., et al. 2004, User's Guide to XMM-Newton Science Analysis System
\bibitem[Favata et al.(2004)]{fav04}Favata, F., Micela, G., Baliunas, S.L. et al. 2004, A\&A, 418, L13
\bibitem[Golub et al.(1982)]{eins}Golub, L., Harnden, F.R.Jr., Pallacicini, R., et al. 1982, ApJ 253,242
\bibitem[Grevesse \& Sauval(1998)]{grev98}Grevesse, N., Sauval, A.J. 1998, Space Sci. Rev., 85, 161
\bibitem[Hempelmann et al.(2003)]{hem03}Hempelmann, A., Schmitt, J.H.M.M., Baliunas, S.L., Donahue, R.A., A\&A, 406, L39
\bibitem[Pagano et al.(2004)]{pag04}Pagano, I., Linsky, J.L., Valenti, J., Duncan, D.K. 2004, A\&A, 415, 331
\bibitem[Robrade et al.(2004)]{eqpeg}Robrade, J., Ness, J.-U., Schmitt, J.H.M.M. 2004, A\&A, 413, 317
\bibitem[Raassen et al.(2003)]{letgs}Raassen, A.J.J., Ness, J.-U., Mewe, R., et al. 2003, A\&A, 400, 671
\bibitem[Schmitt \& Liefke(2004)]{schmitt04}Schmitt, J.H.M.M., Liefke, C. 2004, A\&A, 417, 651
\bibitem[Smith et al.(2001)]{apec}Smith, R.K., Brickhouse, N.S., Liedahl, D.A., Raymond, J.C. 2001, ASP Conf. Series, 247, 161
\end{thebibliography}
\end{document}